\documentclass[pre,aps,twocolumn,showpacs]{revtex4}
\usepackage{graphicx}
\usepackage{color}
\usepackage{epsf}
\usepackage{amsmath}
\usepackage{amssymb}
\usepackage[english]{babel}
\usepackage{bm}
\usepackage{graphics}
\usepackage{color}
\usepackage{dcolumn}
\usepackage{sidecap}
\usepackage{enumitem}

\usepackage{xcolor}
\usepackage{floatrow}

\newcommand {\cc} {\ensuremath{{\bf c}}}
\newcommand {\DD} {\ensuremath{{\bf D}}}

\newcommand {\xxi} {\ensuremath{{\bf \xi}}}

\newcommand{\bes}{ \begin{equation} \begin{split} }
\newcommand{\ees}{ \end{split} \end{equation} }

\newcommand {\PPhi} {\ensuremath{{\bf \Phi}}}
\newcommand {\pphi} {\ensuremath{{\bf \phi}}}

\newcommand{\ignore}[1]{}

\newcommand{\red}[1]{{\color{red}{#1}}}

\begin{document}

\title{Time-varying coupling functions: dynamical inference and\\ cause of synchronization transitions}

\author{Tomislav Stankovski$^{1,2}$}%
\email{t.stankovski@ukim.edu.mk}%

\affiliation{$^1$Faculty of Medicine, Ss Cyril and Methodius University, 50 Divizija 6, Skopje 1000, Macedonia}
\affiliation{$^2$Department of Physics, Lancaster University, Lancaster, LA1 4YB, United Kingdom}

\begin{abstract}
Interactions in nature can be described by their coupling strength, direction of coupling and coupling function. The coupling strength and directionality are relatively  well understood and studied, at least for two interacting systems, however there can be a complexity in the interactions uniquely dependent on the coupling functions. Such a special case is studied here -- synchronization transition occurs only due to the time-variability of the coupling functions, while the net coupling strength is constant throughout the observation time. To motivate the investigation, an example is used to present an analysis of cross-frequency coupling functions between delta and alpha brainwaves extracted from the electroencephalography (EEG) recording of a healthy human subject in a free-running resting state. The results indicate that  time-varying coupling functions are a reality for biological interactions. A model of phase oscillators is used to demonstrate and detect the synchronization transition caused by the varying coupling functions, during an invariant coupling strength. The ability to detect this phenomenon is discussed with the method of dynamical Bayesian inference, which was able to infer the time-varying coupling functions. The form of the coupling function acts as an additional dimension for the interactions and it should be taken into account when detecting biological or other interactions from data.

\end{abstract}


\date{\today}

\maketitle


\section{Introduction}\label{sec1:Intro}

Interacting dynamical systems abound in nature, with examples ranging from physics, biology, climate and social sciences \cite{Pikovsky:01,Strogatz:03b,Haken:83}. Many of them  are open systems and have connections with other systems and the environment. The connectivity presents a link between two dynamical systems, which can be structural, functional or effective \cite{Friston:11}.  Very often these interactions are assessed successfully to a relatively great extend when they are considered isolated, however in some cases there is additional complexity due to external influences and the existing time-variability. From the aspect of mathematical models, such systems and their interactions are studied as non-autonomous dynamical systems \cite{Kloeden:11,Stankovski:13a,Suprunenko:13,Clemson:14b,Lancaster:15}. The time-variability can have different effects on the interactions, including for example, changes in frequency, emergence or disappearance of connectivity, transitions to or out of qualitative states, or time-varying form of the coupling functions.

Coupling function describes in great detail the physical rule of \emph{how} the interactions occur and manifest. For example, the reconstructed coupling function of Belousov-Zhabotinsky chemical interactions revealed how there could be higher-harmonics and bi-stability of the synchronization state \cite{Miyazaki:06}, the knowledge of the coupling function of one pairwise interaction was used to predict the synchronization and clustering of a network of electrochemical oscillators \cite{Kiss:05}, and the form of the cardiorespiratory coupling function was linked to respiratory sinus arrythmia (RSA), a known mechanism in physiology \cite{Iatsenko:13a,Kralemann:13b}.
The coupling function as a whole can be described in terms of its  strength and {\it form}. It is the functional form that has provided a new dimension and perspective, probing directly the functional mechanisms of the interactions. In this way the coupling function can determine the possibility of qualitative transitions between states of the systems e.g.\ routes into and out of synchronization. Decomposition of a coupling function can also facilitate a description of the functional contributions from each separate subsystem within the coupling relationship. Different methods for coupling function detection have been applied widely in chemistry \cite{Kiss:07,Miyazaki:06,Tokuda:07,Kiss:05,Blaha:11}, in cardiorespiratory physiology \cite{Kralemann:13b,Stankovski:12b,Iatsenko:13a}, in neuroscience \cite{Stankovski:15a,Stankovski:16,Wilting:15}, in mechanical interactions \cite{Kralemann:08}, in social sciences \cite{Ranganathan:14} and in secure communications \cite{Stankovski:14a}. The study of coupling function is a very active and expanding field of research \cite{Stankovski:16b}.

\begin{figure*}
\floatbox[{\capbeside\thisfloatsetup{capbesideposition={right,top},capbesidewidth=4.4cm}}]{figure}[\FBwidth]
{\caption{(Color online) An example of time-varying delta-to-alpha neural cross-frequency coupling functions. (a) presents the time-variability of two coupling sub-component strengths $c_1(t)$ and $c_2(t)$. The latter are scaling parameters of the $\sin(\phi_\delta-\phi_\alpha)$ and $\cos(\phi_\delta-\phi_\alpha)$ sub-coupling components in the phase dynamics of the delta-to-alpha interactions. (b) shows three delta-to-alpha coupling functions $q_\alpha(\phi_\delta,\phi_\alpha)$ for three specific time instances as indicated by the grey arrows. Note the variability of the form of the coupling functions for the different time instances. \label{fig1} }}
{\includegraphics[width=0.67\textwidth,angle=0]{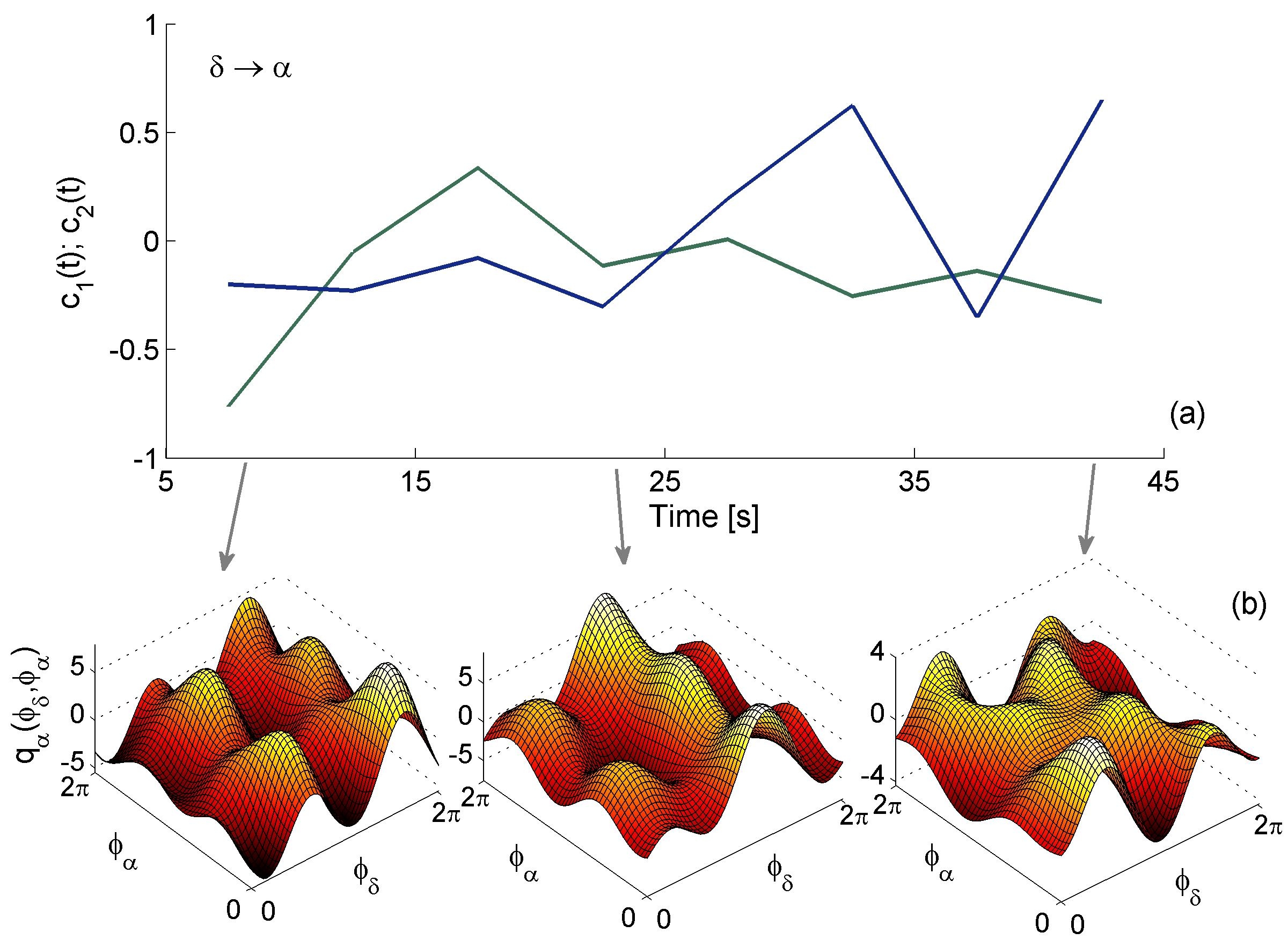}}
\end{figure*}

Recently, it was found that in  interacting biological oscillating systems, not only the frequency and the coupling strength, but also the form of the coupling functions can be a time-varying process \cite{Stankovski:12b}. This was demonstrated in the case of cardiorespiratory interactions and showed that not only the net parametric and quantitative properties, but also the mechanisms of the interactions can be time-varying. These varying coupling functions can change in time the physical rules for the interactions, which can cause transitions of physical effects and phenomena, as the most important outcome of the interactions. For example, the time-varying cardiorespiratory coupling function was shown to induce transitions in and out of synchronization, and between different synchronization ratios \cite{Stankovski:12b}. Therefore, understanding the effects of the time-varying coupling functions on the interactions is of great interest, especially in understanding, detecting and interpreting the interactions of open (biological) systems. In this paper, we further investigate the effects of the time-varying coupling functions on the interactions which can be revealed uniquely by the coupling function analysis.

The time-varying coupling functions are especially important for the detection and inference of interactions from data. Namely, they can be represented by a series of sub-coupling components and their time-variability can introduce a certain complexity in the interactions. Being able to detect and correctly interpret the coupling will depend greatly on the nature of the methods used. For example, some methods assess the amount of information, the net coupling strength and directionality, while other methods perform dynamical inference of a model with differential equations and can describe the underlying mechanisms and coupling functions. Hence, one of the main aspects of the paper will be to discuss a method for analysis of the interactions, in the light of the complexity introduced by the time-varying coupling functions.

\section{Biological interactions as motivation -- the case of neural coupling functions}\label{sec:EEG}

Biological systems are of great importance in this field as they are a classical example of systems that are not isolated but interact with other systems in the body and can be affected by other forces in the local environment. Therefore, the correct analysis of such systems has great implications for the determination and treatment of various physiological states and diseases.

The electrophysiology of the brain evaluated through electroencephalography (EEG) presents an important characteristic for investigating different neural states and diseases, however there could be a certain complexity due to time-varying coupling functions when one tries to analyze the neural interactions associated with this electrophysiology. Here we present the neural cross-frequency coupling functions \cite{Stankovski:15a} as an example of biological systems whose coupling functions are time-varying. For this purpose, the EEG signal of a human subject in a resting state from the public PhysioNet database is used \cite{Goldberger:00,physionetlink,Schalk:04}. The EEG signals were recorded with 64 electrodes according to the standard international 10-10 system. Only \red{one EEG signal recorded at the frontal Fp1 electrode} of the 10-10 system, recorded during eyes-closed resting state of a healthy subject, was analysed here.  After extracting the phases for the delta and alpha brainwave oscillations of the filtered delta and alpha brainwave signals from the EEG signal, the neural cross-frequency coupling functions and the coupling strength of two components are reconstructed by dynamical Bayesian inference (for details about the method see Sec.\ \ref{sec:DBI} below). The two sub-coupling components were chosen arbitrarily for better presentation and in accordance with the other numerical examples that follow.

The delta-to-alpha neural cross-frequency coupling has been found to be generally higher in the eyes-closed than in the eyes-open condition of the resting state,  mostly located within the frontal \red{(e.g.\ at Fp1)} and the parieto-occipital regions, and these regions were connected through larger-scale coupling with a different coupling direction \cite{Jirsa:13}; the delta-alpha coupling was significantly increased with more similar forms of the coupling functions from awake to deep general anaesthesia, and this effect was higher for anaesthesia induced with sevoflurane than with the propofol anaesthetic \cite{Stankovski:16}; and
a strong link between delta and alpha brain activity was found during non-REM sleep, although alpha
waves were greatly diminished and delta waves were dominant \cite{Bashan:12}. Fig.\ \ref{fig1} (a) shows that two of the delta-alpha sub-coupling components have significant time-variability between the allowed $[-1,1]$ values. This leads to the time-variability of the form of the coupling functions  -- as can be seen by comparing the three delta-alpha coupling functions in Fig.\ \ref{fig1} (b). The fact that the sub-coupling components are time-varying, causing the form of the coupling functions to vary as well, points to the possibility of qualitative transitions of the interactions. Therefore, revealing this complexity can be of crucial importance for the correct analysis and interpretation of the neural interactions.

\begin{figure*}
{\caption{(Color online) Time-varying coupling functions lead to synchronization transition, during constant net coupling strength. (a) the time variability of the sub-coupling components $c_1(t)$ and $c_2(t)$ and the constant net coupling strength $\varepsilon_1(t)$. Coupling functions at the beginning (b) and end (c) of the observation time. Note the different form, comparing (b) and (c). Synchronization transition (d) shown with the phase difference $\psi(t)$ (axis right) and the synchronization index $I_{sync}(t)$ (axis left).  \label{fig2} }}
{\includegraphics[width=0.97\textwidth,angle=0]{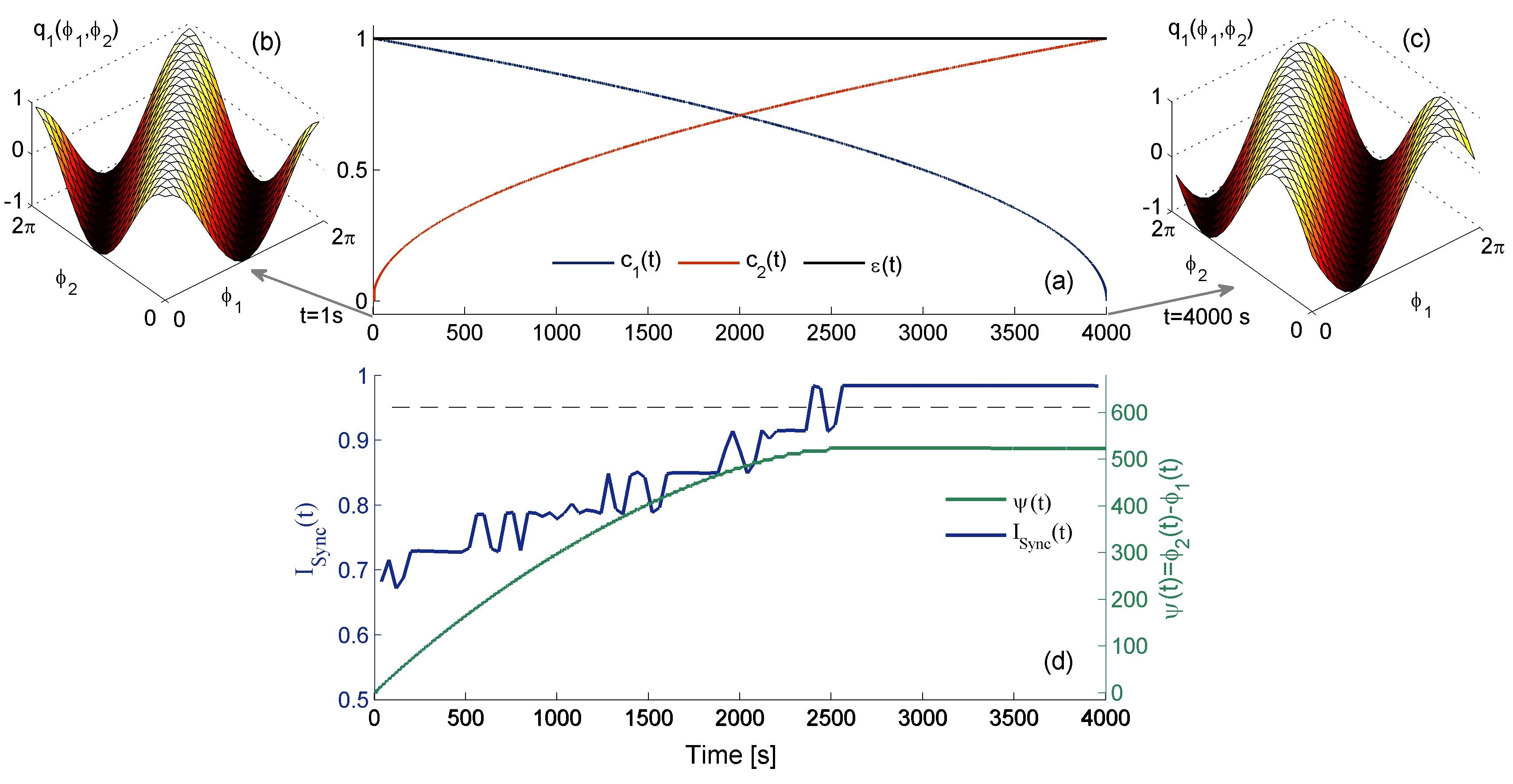}}
\end{figure*}

\section{The Characteristic problem of time-varying coupling functions-- direct approach}

The sub-coupling components define the form of the net function i.e.\ the coupling function is evaluated as a group sum of all the functional components of some set of decomposition functions. For example, for the phase dynamics of interacting oscillators, because of the periodicity, one can decompose the coupling functions into Fourier series. Therefore, the variations of some of the sub-coupling components will also define how the form of a net coupling function will vary. The latter effect, may or may not change the coupling strength, to a greater or lesser extent, however it will change the form of the coupling function and thus the mechanism underlying the interactions, which in turn can cause qualitative transitions. The mechanism is defined by the function that gives the rule through which the input values are translated into output values \cite{Friston:11,Barrett:13}.

To further explain and emphasize the complexity and the effects from the varying sub-coupling components, a simple numerical example is used. A system of two interacting phase oscillators \cite{Kuramoto:84,Petkoski:12a} is considered:
\begin{equation}\label{eq:model}
\begin{split}
   \dot \phi_1&=\omega_1+\varepsilon_1(t) q_1(\phi_1,\phi_2)=\\ &=\omega_1+c_1(t)\sin(\phi_2-\phi_1)+c_2(t)\cos(\phi_2-\phi_1)\\\text{ }\\
   \dot \phi_2&=\omega_2+\varepsilon_2(t) q_2(\phi_1,\phi_2)=\\ &=\omega_2+c_3(t)\sin(\phi_1-\phi_2),
   \end{split}
\end{equation}
where $\omega_1$,$\omega_2$ are the natural frequencies, while $c_1$,$c_2$,$c_3$ are the strengths of the sub-coupling components that scale the appropriate coupling function components. The numerical values of the parameters were set to $\omega_1=2.28$, $\omega_2=1.2$ and $c_3(t)=0.1$. Here, and throughout the manuscript, the focus is  on the coupling in only one direction -- the coupling from the second to the first phase oscillator. In this coupling direction the strengths of the sub-coupling components $c_1(t)$ and $c_2(t)$ are set to be time-varying i.e.\ $c_1(t)=\sqrt{t/T}$ and $c_2(t)=\sqrt{(T-t)/T}$, where the time changes $t=0\rightarrow T$ and $T$ is the total time of observation. In this way $c_1(t)$ varies in the interval $[0,1]$, and $c_2(t)$ varies in the interval $[1,0]$. The total net coupling strength, or the net information flow in one coupling direction, is evaluated as the Euclidian norm of all the coupling components \cite{Kralemann:08}, which in this case is $\varepsilon_1(t)=\sqrt{c_1^2(t)+c_2^2(t)}.$

The main purpose of the example is to present \emph{constant net coupling strength, but time-varying coupling functions.} To do this $c_1(t)$ and $c_2(t)$ were varied in the interval $[0,1]$ simultaneously, but reducing one and increasing the other, in such a way as to give constant net coupling $\varepsilon_1(t)$ for all time. Fig.\ \ref{fig2} (a) shows the time-variation of the coupling parameters, with  $c_1(t)$ and $c_2(t)$ varying, while  $\varepsilon_1(t)$ is being constant. The important part is that even though the net coupling is constant, the coupling function also varies due to the variations of the sub-coupling components.. The latter can be observed by comparing the different forms of the coupling functions at the beginning Fig.\ \ref{fig2} (b) and the end of the observation time Fig.\ \ref{fig2} (c).

The variation of the coupling function changes the mechanism and the physical rule under which the interactions are manifesting, which can lead to qualitative transitions into or out of certain physical effects and phenomena of the interactions. The latter could include transitions to synchronization, amplitude or oscillator death, clustering in networks or the emergence of chimera states. With the current example we present the transition to synchronization. Fig.\ \ref{fig2} (d),(axis right) shows that the phase difference of the interacting system (\ref{eq:model}) is not bounded at the beginning, and then at approximately $t\sim2500s$ there is a transition to bounded phase difference, as in the case of synchronized systems. The same can be verified by the synchronization detection index \cite{Tass:98}, which quantifies, in the interval $[0,1]$, the synchronization (or phase coherence) from the phase difference. The method first divides each phase interval into $N$ bins ($N=10$ used here, and window size 50s). Then, for each bin $l$, it calculates the dependence of $\phi_2(t_j)$ \red{on $\phi_1(t_j)$}, such that $\phi_1(t_j)$ belongs to this bin $l$, and $M_l$ is the number of points in the bin. The average over all bins leads to the final index:
$
I_{Sync}(t_j)=1/N\sum_{l=1}^N |M^{-1}_l\sum e^{i \phi_2(t_j)}|.
$
In this way, $I_{Sync}$ measures the conditional probability for $\phi_2$ to have a certain value provided $\phi_1$ is in a particular bin. If the values are greater than $\sim0.95$ (surrogate level resulting from the mean plus two standard deviations of synchronization indexes from phase random shuffling realizations \cite{Schreiber:00b,Stankovski:14c}, shown with dashed lines in Fig.\ \ref{fig2} (d)), synchronization is detected. Fig.\ \ref{fig2} (d) (axis left) shows consistently that at approximately $t\sim2500s$ there is a transition to synchronization. Therefore, although the net coupling strength was constant between the systems, due to the change of the coupling functions there is a transition to synchronization.

\section{Interaction analysis -- inverse approach}

The main objective here is to be able to detect or infer from data the coupling relations between the systems. This constitutes the so-called inverse approach, starting from the data and attempting to learn the nature of the connectivity between  dynamical systems. The methods for coupling inference have great importance as they allow different \emph{applications} in various fields. 

Many efficient methods for coupling detection have been designed  \cite{Smirnov:09,Andrzejak:06a,Palus:03a,Bahraminasab:08,Jamsek:10,Hlavackova:07,Faes:11,
Stankovski:12b,Kralemann:08,Kiss:05,Tokuda:07}, and they are based around two main aspects. The first aspect is that the assessment of the strength of the interaction and its predominant direction can be used to establish \emph{if} certain interactions exist at all. In this way, one can determine whether some apparent interactions are in fact genuine, and whether the systems under study are really connected or not. The second aspect is that one can infer the underlying interacting systems, including the appropriate coupling functions, and learn about \emph{how} an interaction occurs. The former aspect determines the existence of coupling and relates to functional connectivity, while the latter describes the mechanisms and is part of the effective connectivity methods \cite{Friston:11}. Below, an effective connectivity method which infers coupling functions inherently is presented in light of the interaction complexity, as discussed in the previous sections.

\subsection{Dynamical Bayesian Inference}\label{sec:DBI}

Dynamical inference of coupling functions is a class of effective connectivity methods for coupling assessment.  The main pillar of the procedure is a method for dynamical inference, often referred to as {\it dynamical modelling} or {\it dynamical filtering} \cite{Toussaint:11,Kalman:60,Arulampalam:02,Voss:04,Stankovski:14c}. The dynamical inference procedure starts with the data from two (or more) interacting dynamical systems and uses a method that infers a dynamical model in terms of (ordinary or stochastic) differential equations. The coupling functions are an integral part of, and can be extracted from, the inferred model.

There are a number of other methods for dynamical inference and coupling functions assessment, including those based on least squares and kernel smoothing fits \cite{Rosenblum:01,Kralemann:13b}, dynamical Bayesian inference \cite{Stankovski:12b}, maximum likelihood (multiple-shooting) methods \cite{Tokuda:07}, stochastic modeling \cite{Schwabedal:10} and the phase resetting curve \cite{Levnajic:11}. Below, the dynamical Bayesian inference \cite{Stankovski:12b} will be presented and applied.

\begin{figure*}
\floatbox[{\capbeside\thisfloatsetup{capbesideposition={right,top},capbesidewidth=5cm}}]{figure}[\FBwidth]
{\caption{(Color online) The use of the dynamical Bayesian inference for coupling detection under the time-varying coupling functions condition. (a) the time-variability of the coupling components $c_1(t)$,$c_2(t)$ and the constant total coupling $\varepsilon(t)$ from system Equ.\ \ref{eq:model}. Here the inferred parameters are shown as thick transparent lines above the simulated parameters shown with thin darker lines. (b) the detected coupling functions $q_1(\phi_1,\phi_2)$ shown for four characteristic times as indicated by the arrows. \label{fig4} }}
{\includegraphics[width=0.67\textwidth,angle=0]{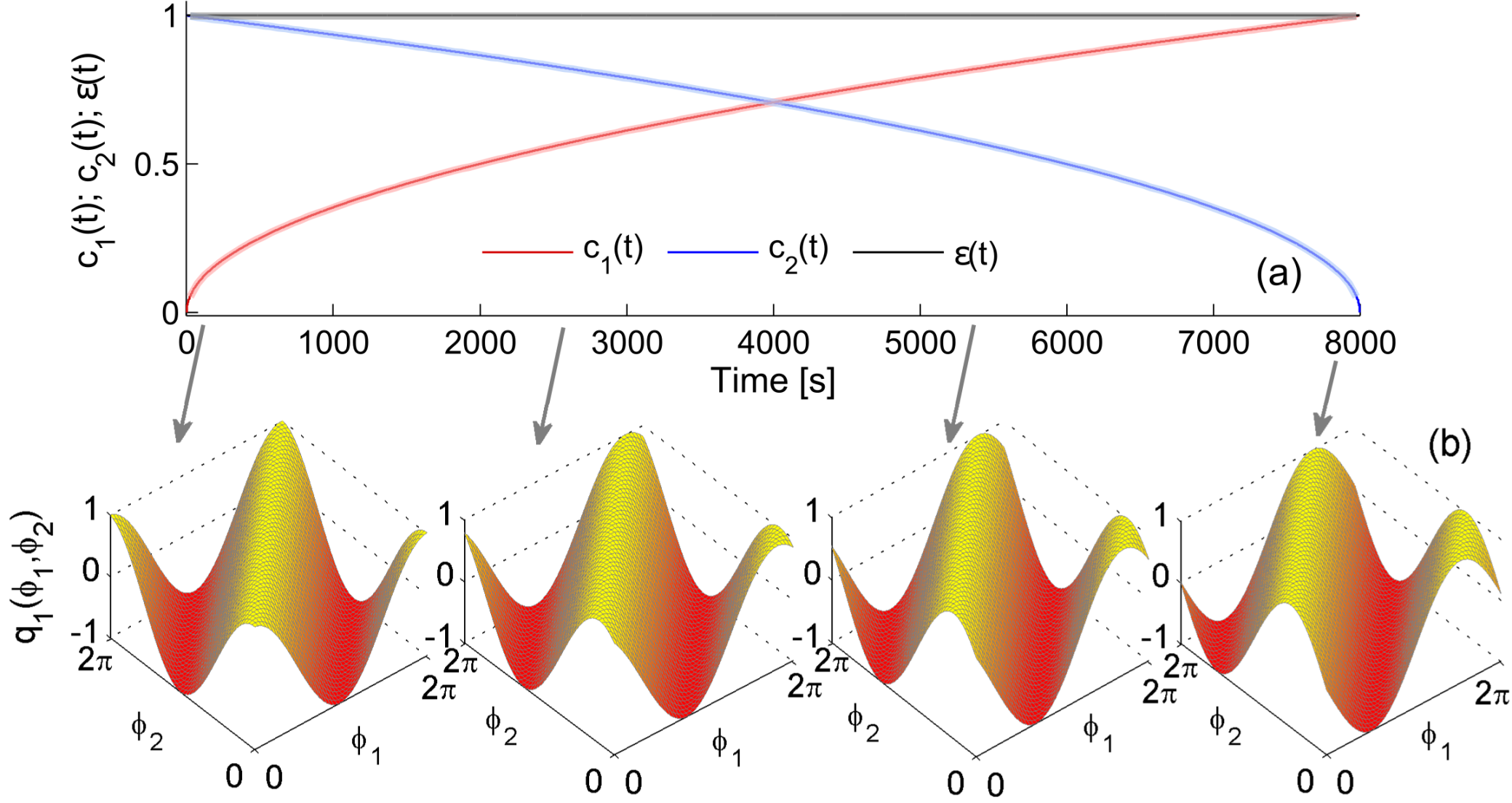}}
\end{figure*}

The signals under consideration are oscillatory and their interactions can be studied effectively through their phase dynamics. Therefore, a model of two coupled phase oscillators \cite{Kuramoto:84} described by the stochastic differential equation is considered:
\begin{equation}
\dot \phi_i= \omega_i  + q_i(\phi_i,\phi_j) + \xi_i(t),
\label{eq:phi}
\end{equation}
with $i\neq j$ for $i,j=\{1,2\}$ and where $\omega_i$ is the parameter for the natural frequency. The deterministic part given by the base functions $q_i(\phi_i,\phi_j)$ describes the self and the interacting dynamics. The external stochastic dynamics $\xi_i(t)$ is considered to be Gaussian white noise $\langle \xi_i(t) \xi_j(\tau)\rangle = \delta(t-\tau) D_{ij}$. Due to the periodic nature of the deterministic dynamics, the base functions can be decomposed into an infinite Fourier series $
q_i(\phi_i,\phi_j)  =    \sum_{s=-\infty}^\infty  \sum_{r=-\infty}^\infty \tilde c_{i;r,s}\, e^{i2\pi r \phi_i}  e^{i2\pi s \phi_j}$. In practice, however, the dynamics are well-described by a finite number of Fourier terms, so that one can rewrite the phase dynamics as:
$$\dot \phi_i=\sum_{k=-K}^{K} \tilde c^{(i)}_k (t) \, \Phi_{i,k}(\phi_i,\phi_j)  + \xxi_i(t),$$
where $\tilde c^{(i)}_0=\omega_i$, and the rest of $\Phi_{i,k}$ and  $\tilde c^{(i)}_k$ are the $K$ most important Fourier components; here we used $K=2$. It is important to note that the net coupling function $q_i(\phi_i,\phi_j)$ is decomposed into a series of sub-coupling components $\Phi_{i,k}(\phi_i,\phi_j)$, which in turn allows for the particular complexity of the interactions to be revealed. In this way, a coupling influence is separated on a large but finite set of sub-functional elements, making it possible to study the coupling contributions of each of them separately or in sub-groups. The assessment of the separate components defines the form of the net coupling function. The Fourier components $\Phi_{i,k}$ act as base functions for the dynamical Bayesian inference, through which the parameters $\tilde{c}_k^{(i)}$ are evaluated.

Dynamical Bayesian inference \cite{Stankovski:12b,Smelyanskiy:05a,Stankovski:15a} enables one to evaluate the model parameters $\tilde \cc$, which give the time-evolving coupling functions and coupling strength in the presence of noise. From Bayes' theorem one can derive the minus log-likelihood function, which is of quadratic form.  Assuming that the parameters are represented as a multivariate normal distribution (with mean $\bar {\cc}$, and covariance matrix ${ {\bf \Sigma} \equiv \Xi^{-1}}$), and given such a distribution for the prior knowledge using the likelihood function, one can calculate recursively \cite{Stankovski:12b,Smelyanskiy:05a} the posterior distribution of the parameters $\tilde \cc_k$ using only the following four equations:
\begin{equation}
\begin{split}
    \label{eq:cD}
     \DD  &= \frac{h}{L} \left(
 \dot{\pphi}_{n} - \cc_k {\PPhi}_{k}({\pphi}_{\cdot,n}^{\ast}) \right)^T \left(\dot{\pphi}_{n} - \cc_k {\PPhi}_{k}({\pphi}_{\cdot,n}^{\ast}) \right) , \\
    {\bf r}_{w}  & = ({\bf \Xi}_\text{prior})_{kw} \,  {\cc}_{w} +
      h \, {\PPhi}_{k}({\pphi}_{\cdot,n}^{\ast}) \,
(\DD^{-1}) \, \dot{{\pphi}}_{n} +\\
 &- \frac{h}{2} \frac{\partial \PPhi_{k}(\phi_{\cdot,n}) }{\partial \pphi}, \\
{\bf \Xi}_{kw}  &= ({{\bf \Xi}_{\text{prior}}})_{kw}   + h \, {\PPhi}_{k}({\pphi}_{\cdot,n}^{\ast}) \,
{(\DD^{-1})} \,   {\PPhi}_{w}({\pphi}_{\cdot,n}^{\ast}),\\
\tilde \cc_k &= ({\bf \Xi}^{-1})_{kw} \,  {\bf r}_{w} ,
\end{split}
\end{equation}
where summation over $n=1,\ldots,N$ is assumed, and summation over repeated indices $k$ and $w$ is implicit. We used informative priors and a special procedure for the propagation of information between consecutive data windows \cite{Stankovski:12b}, which permitted the inference parameters that varied with time (for implementation and usage see \cite{Stankovski:14d}). Once we have the inferred parameters $\tilde \cc$, we can calculate the coupling quantities and characteristics. The coupling functions are evaluated on a $2\pi\times2\pi$ grid using the relevant base functions i.e.\ Fourier components scaled by their inferred coupling parameters. The method has been also generalized and applied to networks of oscillators \cite{Duggento:12,Stankovski:15a}.

The method is applied on data generated from the model of two interacting phase oscillators (Equ.\ \ref{eq:model}), with the same time-variability of the coupling sub-components $c_1(t)$,$c_2(t)$ and constant total net coupling $\varepsilon(t)$ (Fig.\ \ref{fig4} (a)) and the frequency parameters set to $\omega_1=9$ and $\omega_2=1.2$ to ensure that there is no synchronization between the systems.  The latter is important, if there is strong phase synchronization the coupling method will not be able to correctly infer the couplings \cite{Rosenblum:01}.
The inference of the model parameters is very precise and the inferred time-varying coupling sub-components and the net coupling strength are practically indistinguishable from the original simulated parameters -- Fig.\ \ref{fig4} (a). Thus, the dynamical inference provides the information of the whole model and the variations of the sub-function components. The latter define the form of the net coupling function, hence the method is able to follow its time-evolution as well -- compare the coupling functions from left to right in Fig.\ \ref{fig4} (b). As discussed, even though the coupling strength is constant these varying coupling functions can cause qualitative transitions in the interactions.

\section{Discussion and Conclusion}

Much effort has been concentrated in the study of coupling functions with significant progress on theory, methods and application. However, many open questions relevant to the coupling functions still exist. One particular aspect is their assessment, or what can be done once the coupling functions have been determined. The effect of time-varying coupling functions relates and further extends this aspect.

The time-variability of the sub-coupling components and the coupling functions are quite pronounced in  biological systems. This was first observed in cardiorespiratory interactions and demonstrated here with the neural coupling functions of the human resting state (Fig.\ \ref{fig1}). It is important to note that the variability of the biological coupling functions can depend on the physiological state or disease. For example, it was observed that the cardiorespiratory coupling functions are less varying in young compared to old human subjects \cite{Iatsenko:13a,Kralemann:13b}. Similarly, it was found that general anaesthesia reduces the variability of the delta-alpha neural coupling functions \cite{Stankovski:16}. The complexity arising from the varying coupling functions and their analysis is thus of great importance for the assessment of different medical states and diseases.

The theoretical case presented shows the relevance of the time-varying coupling functions in a special situation when the net coupling is invariant and the transition to synchronization is only due to the variations of the form of the coupling functions (Fig.\ \ref{fig2}). The treatment of the same phenomenon with the non-autonomous theory is an interesting open question for future developments. The particular example (Eq.\ \ref{eq:model}) was chosen to be simple and elementary, as it elaborates rather complicated concepts, however the same phenomenon can be investigated also with more complex systems, like limit-cycle oscillators or higher-dimension chaotic systems. Importantly, the phenomenon could have even more complex implications for networks \cite{Barabasi:02} and various methods for networks inference \cite{Timme:07,Levnajic:10,Sysoev:14,Levnaji:14,Timme:14}. Also, the variability is quite simple -- only in two sub-coupling components.  In real systems, the variations may be less pronounced but spread across more sub-coupling components. Needless to say, the situation of having exactly constant coupling strength while having large coupling function variations is quite a special case. In reality it is expected  that the form variations are only partially contributing to the change of the interactions. The coupling strength and the coupling function are strongly related and to some extend dependant, but as shown here they can also affect the outcome of the interactions \emph{independently}. This is because the form of the coupling function acts as an additional dimension in the assessment of the interactions.

The information theoretic based methods assess the statistical dependencies between the signals from the interacting systems. The most prominent methods of this kind are based on Granger causality, transfer entropy,  mutual information and symbolic transfer entropy \cite{Granger:69,Barrett:13,Schreiber:00a,Palus:03a,Staniek:08}. They are statistical measures that can determine the causal relation and the predominant direction of influence, thus measuring a directed functional connectivity. In this way, they usually reveal only the net coupling and direction, thus they are not able to detect the varying sub-coupling components. This is because these methods are designed to perform in such a way i.e.\  designed to infer only the net statistical effects \cite{Barrett:13}. In fact, they are very useful methods for identification of directed connectivity and interaction existence, especially in the case where a model of differential equations for the dynamical inference can not be reconstructed. The presentation in this paper points to the importance of the phenomenon of time-varying sub-coupling components, and hopefully it will \emph{stimulate} further developments of some aspects of these methods. Here, it is worth noting out that there have recently been efforts to use information based methods to perform coupling decomposition \cite{Faes:15,Porta:15} by exploiting the conditional functional dependencies. These methods could advance the detection of certain sub-coupling relations, without exploiting a dynamical model.

The dynamical inference methods are by design able to infer a dynamical model including the coupling functions. This was demonstrated by the use of dynamical Bayesian inference which was able to reveal the sub-coupling functions time-variability (Fig.\ \ref{fig4}). The design and application of dynamical inference methods is rapidly evolving, promising to reveal even more interacting complexities, unique and dependant on the coupling functions. The latter is even more pressing, because as shown the varying coupling functions are a reality for biological systems, and they could have important implications for the interactions in general.

\acknowledgments
The author thanks Tiago Pereira, Martin Rasmussen, Michael Thompson and Rachel Sparks for valuable discussions on the topic, and the Institute of Pathophysiology and Nuclear Medicine, Faculty of Medicine in Skopje, and the Nonlinear and Biomedical Physics group at Lancaster University for the research and financial support.


\begin{thebibliography}{62}
\expandafter\ifx\csname natexlab\endcsname\relax\def\natexlab#1{#1}\fi
\expandafter\ifx\csname bibnamefont\endcsname\relax
  \def\bibnamefont#1{#1}\fi
\expandafter\ifx\csname bibfnamefont\endcsname\relax
  \def\bibfnamefont#1{#1}\fi
\expandafter\ifx\csname citenamefont\endcsname\relax
  \def\citenamefont#1{#1}\fi
\expandafter\ifx\csname url\endcsname\relax
  \def\url#1{\texttt{#1}}\fi
\expandafter\ifx\csname urlprefix\endcsname\relax\def\urlprefix{URL }\fi
\providecommand{\bibinfo}[2]{#2}
\providecommand{\eprint}[2][]{\url{#2}}

\bibitem[{\citenamefont{Pikovsky et~al.}(2001)\citenamefont{Pikovsky,
  Rosenblum, and Kurths}}]{Pikovsky:01}
\bibinfo{author}{\bibfnamefont{A.}~\bibnamefont{Pikovsky}},
  \bibinfo{author}{\bibfnamefont{M.}~\bibnamefont{Rosenblum}},
  \bibnamefont{and} \bibinfo{author}{\bibfnamefont{J.}~\bibnamefont{Kurths}},
  \emph{\bibinfo{title}{Synchronization -- A Universal Concept in Nonlinear
  Sciences}} (\bibinfo{publisher}{Cambridge University Press},
  \bibinfo{address}{Cambridge}, \bibinfo{year}{2001}).

\bibitem[{\citenamefont{Strogatz}(2003)}]{Strogatz:03b}
\bibinfo{author}{\bibfnamefont{S.~H.} \bibnamefont{Strogatz}},
  \emph{\bibinfo{title}{Sync: The Emerging Science of Spontaneous Order}}
  (\bibinfo{publisher}{Hyperion}, \bibinfo{address}{New York},
  \bibinfo{year}{2003}).

\bibitem[{\citenamefont{Haken}(1983)}]{Haken:83}
\bibinfo{author}{\bibfnamefont{H.}~\bibnamefont{Haken}},
  \emph{\bibinfo{title}{Synergetics, An Introduction}}
  (\bibinfo{publisher}{Springer}, \bibinfo{address}{Berlin},
  \bibinfo{year}{1983}).

\bibitem[{\citenamefont{Friston}(2011)}]{Friston:11}
\bibinfo{author}{\bibfnamefont{K.~J.} \bibnamefont{Friston}},
  \bibinfo{journal}{Brain. Connect.} \textbf{\bibinfo{volume}{1}},
  \bibinfo{pages}{13} (\bibinfo{year}{2011}).

\bibitem[{\citenamefont{Kloeden and Rasmussen}(2011)}]{Kloeden:11}
\bibinfo{author}{\bibfnamefont{P.~E.} \bibnamefont{Kloeden}} \bibnamefont{and}
  \bibinfo{author}{\bibfnamefont{M.}~\bibnamefont{Rasmussen}},
  \emph{\bibinfo{title}{Nonautonomous Dynamical Systems}}
  (\bibinfo{publisher}{AMS Mathematical Surveys and Monographs},
  \bibinfo{address}{New York}, \bibinfo{year}{2011}).

\bibitem[{\citenamefont{Stankovski}(2013)}]{Stankovski:13a}
\bibinfo{author}{\bibfnamefont{T.}~\bibnamefont{Stankovski}},
  \emph{\bibinfo{title}{Tackling the Inverse Problem for Non-Autonomous
  Systems: Application to the Life Sciences}} (\bibinfo{publisher}{Springer},
  \bibinfo{address}{Berlin}, \bibinfo{year}{2013}).

\bibitem[{\citenamefont{Suprunenko et~al.}(2013)\citenamefont{Suprunenko,
  Clemson, and Stefanovska}}]{Suprunenko:13}
\bibinfo{author}{\bibfnamefont{Y.~F.} \bibnamefont{Suprunenko}},
  \bibinfo{author}{\bibfnamefont{P.~T.} \bibnamefont{Clemson}},
  \bibnamefont{and}
  \bibinfo{author}{\bibfnamefont{A.}~\bibnamefont{Stefanovska}},
  \bibinfo{journal}{Phys. Rev. Lett.} \textbf{\bibinfo{volume}{111}},
  \bibinfo{pages}{024101} (\bibinfo{year}{2013}).

\bibitem[{\citenamefont{Clemson and Stefanovska}(2014)}]{Clemson:14b}
\bibinfo{author}{\bibfnamefont{P.~T.} \bibnamefont{Clemson}} \bibnamefont{and}
  \bibinfo{author}{\bibfnamefont{A.}~\bibnamefont{Stefanovska}},
  \bibinfo{journal}{Phys. Rep.} \textbf{\bibinfo{volume}{542}},
  \bibinfo{pages}{297} (\bibinfo{year}{2014}).

\bibitem[{\citenamefont{Lancaster et~al.}(2015)\citenamefont{Lancaster,
  Clemson, Suprunenko, Stankovski, and Stefanovska}}]{Lancaster:15}
\bibinfo{author}{\bibfnamefont{G.}~\bibnamefont{Lancaster}},
  \bibinfo{author}{\bibfnamefont{P.~T.} \bibnamefont{Clemson}},
  \bibinfo{author}{\bibfnamefont{Y.~F.} \bibnamefont{Suprunenko}},
  \bibinfo{author}{\bibfnamefont{T.}~\bibnamefont{Stankovski}},
  \bibnamefont{and}
  \bibinfo{author}{\bibfnamefont{A.}~\bibnamefont{Stefanovska}},
  \bibinfo{journal}{Entropy} \textbf{\bibinfo{volume}{17}},
  \bibinfo{pages}{4413} (\bibinfo{year}{2015}).

\bibitem[{\citenamefont{Miyazaki and Kinoshita}(2006)}]{Miyazaki:06}
\bibinfo{author}{\bibfnamefont{J.}~\bibnamefont{Miyazaki}} \bibnamefont{and}
  \bibinfo{author}{\bibfnamefont{S.}~\bibnamefont{Kinoshita}},
  \bibinfo{journal}{Phys. Rev. Lett.} \textbf{\bibinfo{volume}{96}},
  \bibinfo{pages}{194101} (\bibinfo{year}{2006}).

\bibitem[{\citenamefont{Kiss et~al.}(2005)\citenamefont{Kiss, Zhai, and
  Hudson}}]{Kiss:05}
\bibinfo{author}{\bibfnamefont{I.~Z.} \bibnamefont{Kiss}},
  \bibinfo{author}{\bibfnamefont{Y.}~\bibnamefont{Zhai}}, \bibnamefont{and}
  \bibinfo{author}{\bibfnamefont{J.~L.} \bibnamefont{Hudson}},
  \bibinfo{journal}{Phys. Rev. Lett.} \textbf{\bibinfo{volume}{94}},
  \bibinfo{pages}{248301} (\bibinfo{year}{2005}).

\bibitem[{\citenamefont{Iatsenko et~al.}(2013)\citenamefont{Iatsenko, Bernjak,
  Stankovski, Shiogai, Owen-Lynch, Clarkson, McClintock, and
  Stefanovska}}]{Iatsenko:13a}
\bibinfo{author}{\bibfnamefont{D.}~\bibnamefont{Iatsenko}},
  \bibinfo{author}{\bibfnamefont{A.}~\bibnamefont{Bernjak}},
  \bibinfo{author}{\bibfnamefont{T.}~\bibnamefont{Stankovski}},
  \bibinfo{author}{\bibfnamefont{Y.}~\bibnamefont{Shiogai}},
  \bibinfo{author}{\bibfnamefont{P.~J.} \bibnamefont{Owen-Lynch}},
  \bibinfo{author}{\bibfnamefont{P.~B.~M.} \bibnamefont{Clarkson}},
  \bibinfo{author}{\bibfnamefont{P.~V.~E.} \bibnamefont{McClintock}},
  \bibnamefont{and}
  \bibinfo{author}{\bibfnamefont{A.}~\bibnamefont{Stefanovska}},
  \bibinfo{journal}{Phil. Trans. R. Soc. Lond. A}
  \textbf{\bibinfo{volume}{371}}, \bibinfo{pages}{20110622}
  (\bibinfo{year}{2013}).

\bibitem[{\citenamefont{Kralemann et~al.}(2013)\citenamefont{Kralemann,
  Fr{\"u}hwirth, Pikovsky, Rosenblum, Kenner, Schaefer, and
  Moser}}]{Kralemann:13b}
\bibinfo{author}{\bibfnamefont{B.}~\bibnamefont{Kralemann}},
  \bibinfo{author}{\bibfnamefont{M.}~\bibnamefont{Fr{\"u}hwirth}},
  \bibinfo{author}{\bibfnamefont{A.}~\bibnamefont{Pikovsky}},
  \bibinfo{author}{\bibfnamefont{M.}~\bibnamefont{Rosenblum}},
  \bibinfo{author}{\bibfnamefont{T.}~\bibnamefont{Kenner}},
  \bibinfo{author}{\bibfnamefont{J.}~\bibnamefont{Schaefer}}, \bibnamefont{and}
  \bibinfo{author}{\bibfnamefont{M.}~\bibnamefont{Moser}},
  \bibinfo{journal}{Nat. Commun.} \textbf{\bibinfo{volume}{4}},
  \bibinfo{pages}{2418} (\bibinfo{year}{2013}).

\bibitem[{\citenamefont{Kiss et~al.}(2007)\citenamefont{Kiss, Rusin, Kori, and
  Hudson}}]{Kiss:07}
\bibinfo{author}{\bibfnamefont{I.~Z.} \bibnamefont{Kiss}},
  \bibinfo{author}{\bibfnamefont{C.~G.} \bibnamefont{Rusin}},
  \bibinfo{author}{\bibfnamefont{H.}~\bibnamefont{Kori}}, \bibnamefont{and}
  \bibinfo{author}{\bibfnamefont{J.~L.} \bibnamefont{Hudson}},
  \bibinfo{journal}{Science} \textbf{\bibinfo{volume}{316}},
  \bibinfo{pages}{1886} (\bibinfo{year}{2007}).

\bibitem[{\citenamefont{Tokuda et~al.}(2007)\citenamefont{Tokuda, Jain, Kiss,
  and Hudson}}]{Tokuda:07}
\bibinfo{author}{\bibfnamefont{I.~T.} \bibnamefont{Tokuda}},
  \bibinfo{author}{\bibfnamefont{S.}~\bibnamefont{Jain}},
  \bibinfo{author}{\bibfnamefont{I.~Z.} \bibnamefont{Kiss}}, \bibnamefont{and}
  \bibinfo{author}{\bibfnamefont{J.~L.} \bibnamefont{Hudson}},
  \bibinfo{journal}{Phys. Rev. Lett.} \textbf{\bibinfo{volume}{99}},
  \bibinfo{pages}{064101} (\bibinfo{year}{2007}).

\bibitem[{\citenamefont{Blaha et~al.}(2011)\citenamefont{Blaha, Pikovsky,
  Rosenblum, Clark, Rusin, and Hudson}}]{Blaha:11}
\bibinfo{author}{\bibfnamefont{K.~A.} \bibnamefont{Blaha}},
  \bibinfo{author}{\bibfnamefont{A.}~\bibnamefont{Pikovsky}},
  \bibinfo{author}{\bibfnamefont{M.}~\bibnamefont{Rosenblum}},
  \bibinfo{author}{\bibfnamefont{M.~T.} \bibnamefont{Clark}},
  \bibinfo{author}{\bibfnamefont{C.~G.} \bibnamefont{Rusin}}, \bibnamefont{and}
  \bibinfo{author}{\bibfnamefont{J.~L.} \bibnamefont{Hudson}},
  \bibinfo{journal}{Phys.\ Rev.\ E} \textbf{\bibinfo{volume}{84}},
  \bibinfo{pages}{046201} (\bibinfo{year}{2011}).

\bibitem[{\citenamefont{Stankovski et~al.}(2012)\citenamefont{Stankovski,
  Duggento, McClintock, and Stefanovska}}]{Stankovski:12b}
\bibinfo{author}{\bibfnamefont{T.}~\bibnamefont{Stankovski}},
  \bibinfo{author}{\bibfnamefont{A.}~\bibnamefont{Duggento}},
  \bibinfo{author}{\bibfnamefont{P.~V.~E.} \bibnamefont{McClintock}},
  \bibnamefont{and}
  \bibinfo{author}{\bibfnamefont{A.}~\bibnamefont{Stefanovska}},
  \bibinfo{journal}{Phys.\ Rev.\ Lett.} \textbf{\bibinfo{volume}{109}},
  \bibinfo{pages}{024101} (\bibinfo{year}{2012}).

\bibitem[{\citenamefont{Stankovski et~al.}(2015)\citenamefont{Stankovski,
  Ticcinelli, McClintock, and Stefanovska}}]{Stankovski:15a}
\bibinfo{author}{\bibfnamefont{T.}~\bibnamefont{Stankovski}},
  \bibinfo{author}{\bibfnamefont{V.}~\bibnamefont{Ticcinelli}},
  \bibinfo{author}{\bibfnamefont{P.~V.~E.} \bibnamefont{McClintock}},
  \bibnamefont{and}
  \bibinfo{author}{\bibfnamefont{A.}~\bibnamefont{Stefanovska}},
  \bibinfo{journal}{New J.\ Phys.} \textbf{\bibinfo{volume}{17}},
  \bibinfo{pages}{035002} (\bibinfo{year}{2015}).

\bibitem[{\citenamefont{Stankovski
  et~al.}(2016{\natexlab{a}})\citenamefont{Stankovski, Petkoski, Raeder, Smith,
  McClintock, and Stefanovska}}]{Stankovski:16}
\bibinfo{author}{\bibfnamefont{T.}~\bibnamefont{Stankovski}},
  \bibinfo{author}{\bibfnamefont{S.}~\bibnamefont{Petkoski}},
  \bibinfo{author}{\bibfnamefont{J.}~\bibnamefont{Raeder}},
  \bibinfo{author}{\bibfnamefont{A.~F.} \bibnamefont{Smith}},
  \bibinfo{author}{\bibfnamefont{P.~V.~E.} \bibnamefont{McClintock}},
  \bibnamefont{and}
  \bibinfo{author}{\bibfnamefont{A.}~\bibnamefont{Stefanovska}},
  \bibinfo{journal}{Phil. Trans. R. Soc. A} \textbf{\bibinfo{volume}{374}},
  \bibinfo{pages}{20150186} (\bibinfo{year}{2016}{\natexlab{a}}).

  \bibitem[{\citenamefont{Wilting et~al.}(2015)\citenamefont{Wilting and Lehnertz}}]{Wilting:15}
\bibinfo{author}{\bibfnamefont{J.}~\bibnamefont{Wilting}},
  \bibnamefont{and}
  \bibinfo{author}{\bibfnamefont{K.}~\bibnamefont{Lehnertz}},
  \bibinfo{journal}{Eur.\ Phys.\ J B} \textbf{\bibinfo{volume}{88}},
  \bibinfo{pages}{193} (\bibinfo{year}{2015}).

\bibitem[{\citenamefont{Kralemann et~al.}(2008)\citenamefont{Kralemann,
  Cimponeriu, Rosenblum, Pikovsky, and Mrowka}}]{Kralemann:08}
\bibinfo{author}{\bibfnamefont{B.}~\bibnamefont{Kralemann}},
  \bibinfo{author}{\bibfnamefont{L.}~\bibnamefont{Cimponeriu}},
  \bibinfo{author}{\bibfnamefont{M.}~\bibnamefont{Rosenblum}},
  \bibinfo{author}{\bibfnamefont{A.}~\bibnamefont{Pikovsky}}, \bibnamefont{and}
  \bibinfo{author}{\bibfnamefont{R.}~\bibnamefont{Mrowka}},
  \bibinfo{journal}{Phys. Rev. E} \textbf{\bibinfo{volume}{77}},
  \bibinfo{pages}{{066205}} (\bibinfo{year}{2008}).

\bibitem[{\citenamefont{Ranganathan et~al.}(2014)\citenamefont{Ranganathan,
  Spaiser, Mann, and Sumpter}}]{Ranganathan:14}
\bibinfo{author}{\bibfnamefont{S.}~\bibnamefont{Ranganathan}},
  \bibinfo{author}{\bibfnamefont{V.}~\bibnamefont{Spaiser}},
  \bibinfo{author}{\bibfnamefont{R.~P.} \bibnamefont{Mann}}, \bibnamefont{and}
  \bibinfo{author}{\bibfnamefont{D.~J.~T.} \bibnamefont{Sumpter}},
  \bibinfo{journal}{PloS one} \textbf{\bibinfo{volume}{9}},
  \bibinfo{pages}{e86468} (\bibinfo{year}{2014}).

\bibitem[{\citenamefont{Stankovski
  et~al.}(2014{\natexlab{a}})\citenamefont{Stankovski, McClintock, and
  Stefanovska}}]{Stankovski:14a}
\bibinfo{author}{\bibfnamefont{T.}~\bibnamefont{Stankovski}},
  \bibinfo{author}{\bibfnamefont{P.~V.~E.} \bibnamefont{McClintock}},
  \bibnamefont{and}
  \bibinfo{author}{\bibfnamefont{A.}~\bibnamefont{Stefanovska}},
  \bibinfo{journal}{Phys.\ Rev.\ X} \textbf{\bibinfo{volume}{4}},
  \bibinfo{pages}{011026} (\bibinfo{year}{2014}{\natexlab{a}}).

\bibitem[{\citenamefont{Stankovski
  et~al.}(2016{\natexlab{b}})\citenamefont{Stankovski, Pereira, McClintock, and
  Stefanovska}}]{Stankovski:16b}
\bibinfo{author}{\bibfnamefont{T.}~\bibnamefont{Stankovski}},
  \bibinfo{author}{\bibfnamefont{T.}~\bibnamefont{Pereira}},
  \bibinfo{author}{\bibfnamefont{P.~V.~E.} \bibnamefont{McClintock}},
  \bibnamefont{and}
  \bibinfo{author}{\bibfnamefont{A.}~\bibnamefont{Stefanovska}},
  \bibinfo{journal}{Rev. Mod. Phys. -- in submission}
  (\bibinfo{year}{2017}{\natexlab{b}}).

\bibitem[{\citenamefont{Goldberger et~al.}(2000)\citenamefont{Goldberger,
  Amaral, Glass, Hausdorff, Ivanov, Mark, Mietus, Moody, Peng, and
  Stanley}}]{Goldberger:00}
\bibinfo{author}{\bibfnamefont{A.~L.} \bibnamefont{Goldberger}},
  \bibinfo{author}{\bibfnamefont{L.~A.} \bibnamefont{Amaral}},
  \bibinfo{author}{\bibfnamefont{L.}~\bibnamefont{Glass}},
  \bibinfo{author}{\bibfnamefont{J.~M.} \bibnamefont{Hausdorff}},
  \bibinfo{author}{\bibfnamefont{P.~C.} \bibnamefont{Ivanov}},
  \bibinfo{author}{\bibfnamefont{R.~G.} \bibnamefont{Mark}},
  \bibinfo{author}{\bibfnamefont{J.~E.} \bibnamefont{Mietus}},
  \bibinfo{author}{\bibfnamefont{G.~B.} \bibnamefont{Moody}},
  \bibinfo{author}{\bibfnamefont{C.-K.} \bibnamefont{Peng}}, \bibnamefont{and}
  \bibinfo{author}{\bibfnamefont{H.~E.} \bibnamefont{Stanley}},
  \bibinfo{journal}{Circulation} \textbf{\bibinfo{volume}{101}},
  \bibinfo{pages}{e215} (\bibinfo{year}{2000}).

\bibitem[{phy($ $)}]{physionetlink}
\bibinfo{journal}{www.PhysioNet.org}  \bibinfo{year}{$ $}

\bibitem[{\citenamefont{Schalk et~al.}(2004)\citenamefont{Schalk, McFarland,
  Hinterberger, Birbaumer, and Wolpaw}}]{Schalk:04}
\bibinfo{author}{\bibfnamefont{G.}~\bibnamefont{Schalk}},
  \bibinfo{author}{\bibfnamefont{D.~J.} \bibnamefont{McFarland}},
  \bibinfo{author}{\bibfnamefont{T.}~\bibnamefont{Hinterberger}},
  \bibinfo{author}{\bibfnamefont{N.}~\bibnamefont{Birbaumer}},
  \bibnamefont{and} \bibinfo{author}{\bibfnamefont{J.~R.}
  \bibnamefont{Wolpaw}}, \bibinfo{journal}{IEEE Trans. Biomed. Eng.}
  \textbf{\bibinfo{volume}{51}}, \bibinfo{pages}{1034} (\bibinfo{year}{2004}).

\bibitem[{\citenamefont{Jirsa and M{\"u}ller}(2013)}]{Jirsa:13}
\bibinfo{author}{\bibfnamefont{V.}~\bibnamefont{Jirsa}} \bibnamefont{and}
  \bibinfo{author}{\bibfnamefont{V.}~\bibnamefont{M{\"u}ller}},
  \bibinfo{journal}{Frontiers Comput. Neurosci.} \textbf{\bibinfo{volume}{7}},
  \bibinfo{pages}{78} (\bibinfo{year}{2013}).

\bibitem[{\citenamefont{Bashan et~al.}(2012)\citenamefont{Bashan, Bartsch,
  Kantelhardt, Havlin, and Ivanov}}]{Bashan:12}
\bibinfo{author}{\bibfnamefont{A.}~\bibnamefont{Bashan}},
  \bibinfo{author}{\bibfnamefont{R.~P.} \bibnamefont{Bartsch}},
  \bibinfo{author}{\bibfnamefont{J.~W.} \bibnamefont{Kantelhardt}},
  \bibinfo{author}{\bibfnamefont{S.}~\bibnamefont{Havlin}}, \bibnamefont{and}
  \bibinfo{author}{\bibfnamefont{P.~C.} \bibnamefont{Ivanov}},
  \bibinfo{journal}{Nat. Commun.} \textbf{\bibinfo{volume}{3}},
  \bibinfo{pages}{702} (\bibinfo{year}{2012}).

\bibitem[{\citenamefont{Barrett and Barnett}(2013)}]{Barrett:13}
\bibinfo{author}{\bibfnamefont{A.~B.} \bibnamefont{Barrett}} \bibnamefont{and}
  \bibinfo{author}{\bibfnamefont{L.}~\bibnamefont{Barnett}},
  \bibinfo{journal}{Front. Neuroinf.} \textbf{\bibinfo{volume}{7}}
  (\bibinfo{year}{2013}).

\bibitem[{\citenamefont{Kuramoto}(1984)}]{Kuramoto:84}
\bibinfo{author}{\bibfnamefont{Y.}~\bibnamefont{Kuramoto}},
  \emph{\bibinfo{title}{Chemical Oscillations, Waves, and Turbulence}}
  (\bibinfo{publisher}{Springer-Verlag}, \bibinfo{address}{Berlin},
  \bibinfo{year}{1984}).

\bibitem[{\citenamefont{Petkoski and Stefanovska}(2012)}]{Petkoski:12a}
\bibinfo{author}{\bibfnamefont{S.}~\bibnamefont{Petkoski}} \bibnamefont{and}
  \bibinfo{author}{\bibfnamefont{A.}~\bibnamefont{Stefanovska}},
  \bibinfo{journal}{Phys. Rev. E} \textbf{\bibinfo{volume}{86}},
  \bibinfo{pages}{046212} (\bibinfo{year}{2012}).

\bibitem[{\citenamefont{Tass et~al.}(1998)\citenamefont{Tass, Rosenblum, Weule,
  Kurths, Pikovsky, Volkmann, Schnitzler, and Freund}}]{Tass:98}
\bibinfo{author}{\bibfnamefont{P.}~\bibnamefont{Tass}},
  \bibinfo{author}{\bibfnamefont{M.~G.} \bibnamefont{Rosenblum}},
  \bibinfo{author}{\bibfnamefont{J.}~\bibnamefont{Weule}},
  \bibinfo{author}{\bibfnamefont{J.}~\bibnamefont{Kurths}},
  \bibinfo{author}{\bibfnamefont{A.}~\bibnamefont{Pikovsky}},
  \bibinfo{author}{\bibfnamefont{J.}~\bibnamefont{Volkmann}},
  \bibinfo{author}{\bibfnamefont{A.}~\bibnamefont{Schnitzler}},
  \bibnamefont{and} \bibinfo{author}{\bibfnamefont{H.-J.}
  \bibnamefont{Freund}}, \bibinfo{journal}{Phys. Rev. Lett.}
  \textbf{\bibinfo{volume}{81}}, \bibinfo{pages}{3291} (\bibinfo{year}{1998}).

\bibitem[{\citenamefont{Schreiber and Schmitz}(2000)}]{Schreiber:00b}
\bibinfo{author}{\bibfnamefont{T.}~\bibnamefont{Schreiber}} \bibnamefont{and}
  \bibinfo{author}{\bibfnamefont{A.}~\bibnamefont{Schmitz}},
  \bibinfo{journal}{Physica D} \textbf{\bibinfo{volume}{142}},
  \bibinfo{pages}{346} (\bibinfo{year}{2000}).

\bibitem[{\citenamefont{Stankovski
  et~al.}(2014{\natexlab{b}})\citenamefont{Stankovski, McClintock, and
  Stefanovska}}]{Stankovski:14c}
\bibinfo{author}{\bibfnamefont{T.}~\bibnamefont{Stankovski}},
  \bibinfo{author}{\bibfnamefont{P.~V.~E.} \bibnamefont{McClintock}},
  \bibnamefont{and}
  \bibinfo{author}{\bibfnamefont{A.}~\bibnamefont{Stefanovska}},
  \bibinfo{journal}{Phys.\ Rev.\ E} \textbf{\bibinfo{volume}{89}},
  \bibinfo{pages}{062909} (\bibinfo{year}{2014}{\natexlab{b}}).

\bibitem[{\citenamefont{Smirnov and Bezruchko}(2009)}]{Smirnov:09}
\bibinfo{author}{\bibfnamefont{D.~A.} \bibnamefont{Smirnov}} \bibnamefont{and}
  \bibinfo{author}{\bibfnamefont{B.~P.} \bibnamefont{Bezruchko}},
  \bibinfo{journal}{Phys. Rev. E} \textbf{\bibinfo{volume}{79}},
  \bibinfo{pages}{046204} (\bibinfo{year}{2009}).

\bibitem[{\citenamefont{Andrzejak et~al.}(2006)\citenamefont{Andrzejak,
  Ledberg, and Deco}}]{Andrzejak:06a}
\bibinfo{author}{\bibfnamefont{R.}~\bibnamefont{Andrzejak}},
  \bibinfo{author}{\bibfnamefont{A.}~\bibnamefont{Ledberg}}, \bibnamefont{and}
  \bibinfo{author}{\bibfnamefont{G.}~\bibnamefont{Deco}}, \bibinfo{journal}{New
  Journal of Physics} \textbf{\bibinfo{volume}{8}}, \bibinfo{pages}{6}
  (\bibinfo{year}{2006}).

\bibitem[{\citenamefont{Palu{\v{s}} and Stefanovska}(2003)}]{Palus:03a}
\bibinfo{author}{\bibfnamefont{M.}~\bibnamefont{Palu{\v{s}}}} \bibnamefont{and}
  \bibinfo{author}{\bibfnamefont{A.}~\bibnamefont{Stefanovska}},
  \bibinfo{journal}{Phys.\ Rev.\ E} \textbf{\bibinfo{volume}{67}},
  \bibinfo{pages}{055201(R)} (\bibinfo{year}{2003}).

\bibitem[{\citenamefont{Bahraminasab et~al.}(2008)\citenamefont{Bahraminasab,
  Ghasemi, Stefanovska, McClintock, and Kantz}}]{Bahraminasab:08}
\bibinfo{author}{\bibfnamefont{A.}~\bibnamefont{Bahraminasab}},
  \bibinfo{author}{\bibfnamefont{F.}~\bibnamefont{Ghasemi}},
  \bibinfo{author}{\bibfnamefont{A.}~\bibnamefont{Stefanovska}},
  \bibinfo{author}{\bibfnamefont{P.~V.~E.} \bibnamefont{McClintock}},
  \bibnamefont{and} \bibinfo{author}{\bibfnamefont{H.}~\bibnamefont{Kantz}},
  \bibinfo{journal}{Phys. Rev. Lett.} \textbf{\bibinfo{volume}{100}},
  \bibinfo{pages}{084101} (\bibinfo{year}{2008}).

\bibitem[{\citenamefont{Jam\v{s}ek et~al.}(2010)\citenamefont{Jam\v{s}ek,
  Palu\v{s}, and Stefanovska}}]{Jamsek:10}
\bibinfo{author}{\bibfnamefont{J.}~\bibnamefont{Jam\v{s}ek}},
  \bibinfo{author}{\bibfnamefont{M.}~\bibnamefont{Palu\v{s}}},
  \bibnamefont{and}
  \bibinfo{author}{\bibfnamefont{A.}~\bibnamefont{Stefanovska}},
  \bibinfo{journal}{Phys. Rev. E} \textbf{\bibinfo{volume}{81}},
  \bibinfo{pages}{036207} (\bibinfo{year}{2010}).

\bibitem[{\citenamefont{{Hlav\'{a}\v{c}kov\'{a}\'{a}-Schindler}
  et~al.}(2007)\citenamefont{{Hlav\'{a}\v{c}kov\'{a}\'{a}-Schindler},
  Palu\v{s}, Vejmelka, and Bhattacharya}}]{Hlavackova:07}
\bibinfo{author}{\bibfnamefont{K.}~\bibnamefont{{Hlav\'{a}\v{c}kov\'{a}\'{a}-Schindler}}},
  \bibinfo{author}{\bibfnamefont{M.}~\bibnamefont{Palu\v{s}}},
  \bibinfo{author}{\bibfnamefont{M.}~\bibnamefont{Vejmelka}}, \bibnamefont{and}
  \bibinfo{author}{\bibfnamefont{J.}~\bibnamefont{Bhattacharya}},
  \bibinfo{journal}{Phys.\ Rep.} \textbf{\bibinfo{volume}{441}},
  \bibinfo{pages}{1} (\bibinfo{year}{2007}).

\bibitem[{\citenamefont{Faes et~al.}(2011)\citenamefont{Faes, Nollo, and
  Porta}}]{Faes:11}
\bibinfo{author}{\bibfnamefont{L.}~\bibnamefont{Faes}},
  \bibinfo{author}{\bibfnamefont{G.}~\bibnamefont{Nollo}}, \bibnamefont{and}
  \bibinfo{author}{\bibfnamefont{A.}~\bibnamefont{Porta}},
  \bibinfo{journal}{Phys. Rev. E} \textbf{\bibinfo{volume}{83}},
  \bibinfo{pages}{051112} (\bibinfo{year}{2011}).

\bibitem[{\citenamefont{{von Toussaint}}(2011)}]{Toussaint:11}
\bibinfo{author}{\bibfnamefont{U.}~\bibnamefont{{von Toussaint}}},
  \bibinfo{journal}{Rev. Mod. Phys.} \textbf{\bibinfo{volume}{83}},
  \bibinfo{pages}{943} (\bibinfo{year}{2011}).

\bibitem[{\citenamefont{Kalman}(1960)}]{Kalman:60}
\bibinfo{author}{\bibfnamefont{R.~E.} \bibnamefont{Kalman}},
  \bibinfo{journal}{J. Fluid. Eng.} \textbf{\bibinfo{volume}{82}},
  \bibinfo{pages}{35} (\bibinfo{year}{1960}).

\bibitem[{\citenamefont{Sanjeev~Arulampalam
  et~al.}(2002)\citenamefont{Sanjeev~Arulampalam, Maskell, Gordon, and
  Clapp}}]{Arulampalam:02}
\bibinfo{author}{\bibfnamefont{M.}~\bibnamefont{Sanjeev~Arulampalam}},
  \bibinfo{author}{\bibfnamefont{S.}~\bibnamefont{Maskell}},
  \bibinfo{author}{\bibfnamefont{N.}~\bibnamefont{Gordon}}, \bibnamefont{and}
  \bibinfo{author}{\bibfnamefont{T.}~\bibnamefont{Clapp}},
  \bibinfo{journal}{Signal Processing, IEEE Transactions on}
  \textbf{\bibinfo{volume}{50}}, \bibinfo{pages}{174} (\bibinfo{year}{2002}).

\bibitem[{\citenamefont{Voss et~al.}(2004)\citenamefont{Voss, Timmer, and
  Kurths}}]{Voss:04}
\bibinfo{author}{\bibfnamefont{H.~U.} \bibnamefont{Voss}},
  \bibinfo{author}{\bibfnamefont{J.}~\bibnamefont{Timmer}}, \bibnamefont{and}
  \bibinfo{author}{\bibfnamefont{J.}~\bibnamefont{Kurths}},
  \bibinfo{journal}{Int. J. Bifurcat. Chaos} \textbf{\bibinfo{volume}{14}},
  \bibinfo{pages}{1905} (\bibinfo{year}{2004}).

\bibitem[{\citenamefont{Rosenblum and Pikovsky}(2001)}]{Rosenblum:01}
\bibinfo{author}{\bibfnamefont{M.~G.} \bibnamefont{Rosenblum}}
  \bibnamefont{and} \bibinfo{author}{\bibfnamefont{A.~S.}
  \bibnamefont{Pikovsky}}, \bibinfo{journal}{Phys. Rev. E.}
  \textbf{\bibinfo{volume}{64}}, \bibinfo{pages}{045202}
  (\bibinfo{year}{2001}).

\bibitem[{\citenamefont{Schwabedal and Pikovsky}(2010)}]{Schwabedal:10}
\bibinfo{author}{\bibfnamefont{J.~T.} \bibnamefont{Schwabedal}}
  \bibnamefont{and} \bibinfo{author}{\bibfnamefont{A.}~\bibnamefont{Pikovsky}},
  \bibinfo{journal}{Phys. Rev. E} \textbf{\bibinfo{volume}{81}},
  \bibinfo{pages}{046218} (\bibinfo{year}{2010}).

\bibitem[{\citenamefont{Levnaji\'{c} and Pikovsky}(2011)}]{Levnajic:11}
\bibinfo{author}{\bibfnamefont{Z.}~\bibnamefont{Levnaji\'{c}}}
  \bibnamefont{and} \bibinfo{author}{\bibfnamefont{A.}~\bibnamefont{Pikovsky}},
  \bibinfo{journal}{Phys. Rev. Lett.} \textbf{\bibinfo{volume}{107}},
  \bibinfo{pages}{034101} (\bibinfo{year}{2011}).

\bibitem[{\citenamefont{Smelyanskiy et~al.}(2005)\citenamefont{Smelyanskiy,
  Luchinsky, Stefanovska, and McClintock}}]{Smelyanskiy:05a}
\bibinfo{author}{\bibfnamefont{V.~N.} \bibnamefont{Smelyanskiy}},
  \bibinfo{author}{\bibfnamefont{D.~G.} \bibnamefont{Luchinsky}},
  \bibinfo{author}{\bibfnamefont{A.}~\bibnamefont{Stefanovska}},
  \bibnamefont{and} \bibinfo{author}{\bibfnamefont{P.~V.~E.}
  \bibnamefont{McClintock}}, \bibinfo{journal}{Phys.\ Rev.\ Lett.}
  \textbf{\bibinfo{volume}{94}}, \bibinfo{pages}{098101}
  (\bibinfo{year}{2005}).

\bibitem[{\citenamefont{Stankovski
  et~al.}(2014{\natexlab{c}})\citenamefont{Stankovski, Duggento, McClintock,
  and Stefanovska}}]{Stankovski:14d}
\bibinfo{author}{\bibfnamefont{T.}~\bibnamefont{Stankovski}},
  \bibinfo{author}{\bibfnamefont{A.}~\bibnamefont{Duggento}},
  \bibinfo{author}{\bibfnamefont{P.~V.~E.} \bibnamefont{McClintock}},
  \bibnamefont{and}
  \bibinfo{author}{\bibfnamefont{A.}~\bibnamefont{Stefanovska}},
  \bibinfo{journal}{Eur.\ Phys.\ J.\ Special Topics}
  \textbf{\bibinfo{volume}{223}}, \bibinfo{pages}{2685}
  (\bibinfo{year}{2014}{\natexlab{c}}).

\bibitem[{\citenamefont{Duggento et~al.}(2012)\citenamefont{Duggento,
  Stankovski, McClintock, and Stefanovska}}]{Duggento:12}
\bibinfo{author}{\bibfnamefont{A.}~\bibnamefont{Duggento}},
  \bibinfo{author}{\bibfnamefont{T.}~\bibnamefont{Stankovski}},
  \bibinfo{author}{\bibfnamefont{P.~V.~E.} \bibnamefont{McClintock}},
  \bibnamefont{and}
  \bibinfo{author}{\bibfnamefont{A.}~\bibnamefont{Stefanovska}},
  \bibinfo{journal}{Phys. Rev. E} \textbf{\bibinfo{volume}{86}},
  \bibinfo{pages}{061126} (\bibinfo{year}{2012}).

\bibitem[{\citenamefont{Albert and Barab{\'a}si}(2002)}]{Barabasi:02}
\bibinfo{author}{\bibfnamefont{R.}~\bibnamefont{Albert}} \bibnamefont{and}
  \bibinfo{author}{\bibfnamefont{A.-L.} \bibnamefont{Barab{\'a}si}},
  \bibinfo{journal}{Rev. Mod. Phys.} \textbf{\bibinfo{volume}{74}},
  \bibinfo{pages}{47} (\bibinfo{year}{2002}).

\bibitem[{\citenamefont{Timme}(2007)}]{Timme:07}
\bibinfo{author}{\bibfnamefont{M.}~\bibnamefont{Timme}},
  \bibinfo{journal}{Physical review letters} \textbf{\bibinfo{volume}{98}},
  \bibinfo{pages}{224101} (\bibinfo{year}{2007}).

\bibitem[{\citenamefont{Levnaji{\'c} and Pikovsky}(2010)}]{Levnajic:10}
\bibinfo{author}{\bibfnamefont{Z.}~\bibnamefont{Levnaji{\'c}}}
  \bibnamefont{and} \bibinfo{author}{\bibfnamefont{A.}~\bibnamefont{Pikovsky}},
  \bibinfo{journal}{Physical Review E} \textbf{\bibinfo{volume}{82}},
  \bibinfo{pages}{056202} (\bibinfo{year}{2010}).

\bibitem[{\citenamefont{Sysoev et~al.}(2014)\citenamefont{Sysoev, Prokhorov,
  Ponomarenko, and Bezruchko}}]{Sysoev:14}
\bibinfo{author}{\bibfnamefont{I.}~\bibnamefont{Sysoev}},
  \bibinfo{author}{\bibfnamefont{M.}~\bibnamefont{Prokhorov}},
  \bibinfo{author}{\bibfnamefont{V.}~\bibnamefont{Ponomarenko}},
  \bibnamefont{and}
  \bibinfo{author}{\bibfnamefont{B.}~\bibnamefont{Bezruchko}},
  \bibinfo{journal}{Physical Review E} \textbf{\bibinfo{volume}{89}},
  \bibinfo{pages}{062911} (\bibinfo{year}{2014}).

\bibitem[{\citenamefont{Levnaji and Pikovsky}(2014)}]{Levnaji:14}
\bibinfo{author}{\bibfnamefont{Z.}~\bibnamefont{Levnaji}} \bibnamefont{and}
  \bibinfo{author}{\bibfnamefont{A.}~\bibnamefont{Pikovsky}},
  \bibinfo{journal}{Scientific reports} \textbf{\bibinfo{volume}{4}}
  (\bibinfo{year}{2014}).

\bibitem[{\citenamefont{Timme and Casadiego}(2014)}]{Timme:14}
\bibinfo{author}{\bibfnamefont{M.}~\bibnamefont{Timme}} \bibnamefont{and}
  \bibinfo{author}{\bibfnamefont{J.}~\bibnamefont{Casadiego}},
  \bibinfo{journal}{Journal of Physics A: Mathematical and Theoretical}
  \textbf{\bibinfo{volume}{47}}, \bibinfo{pages}{343001}
  (\bibinfo{year}{2014}).

\bibitem[{\citenamefont{Granger}(1969)}]{Granger:69}
\bibinfo{author}{\bibfnamefont{C.~W.} \bibnamefont{Granger}},
  \bibinfo{journal}{Econ: J. Econ. Soc.} pp. \bibinfo{pages}{424--438}
  (\bibinfo{year}{1969}).

\bibitem[{\citenamefont{Schreiber}(2000)}]{Schreiber:00a}
\bibinfo{author}{\bibfnamefont{T.}~\bibnamefont{Schreiber}},
  \bibinfo{journal}{Phys. Rev. Lett.} \textbf{\bibinfo{volume}{85}},
  \bibinfo{pages}{461} (\bibinfo{year}{2000}).

\bibitem[{\citenamefont{Staniek and Lehnertz}(2008)}]{Staniek:08}
\bibinfo{author}{\bibfnamefont{M.}~\bibnamefont{Staniek}} \bibnamefont{and}
  \bibinfo{author}{\bibfnamefont{K.}~\bibnamefont{Lehnertz}},
  \bibinfo{journal}{Phys. Rev. Lett.} \textbf{\bibinfo{volume}{100}},
  \bibinfo{pages}{158101} (\bibinfo{year}{2008}).

\bibitem[{\citenamefont{Faes et~al.}(2015)\citenamefont{Faes, Porta, and
  Nollo}}]{Faes:15}
\bibinfo{author}{\bibfnamefont{L.}~\bibnamefont{Faes}},
  \bibinfo{author}{\bibfnamefont{A.}~\bibnamefont{Porta}}, \bibnamefont{and}
  \bibinfo{author}{\bibfnamefont{G.}~\bibnamefont{Nollo}},
  \bibinfo{journal}{Entropy} \textbf{\bibinfo{volume}{17}},
  \bibinfo{pages}{277} (\bibinfo{year}{2015}).

\bibitem[{\citenamefont{Porta et~al.}(2015)\citenamefont{Porta, Faes, Marchi,
  Bari, De~Maria, Guzzetti, Colombo, and Raimondi}}]{Porta:15}
\bibinfo{author}{\bibfnamefont{A.}~\bibnamefont{Porta}},
  \bibinfo{author}{\bibfnamefont{L.}~\bibnamefont{Faes}},
  \bibinfo{author}{\bibfnamefont{A.}~\bibnamefont{Marchi}},
  \bibinfo{author}{\bibfnamefont{V.}~\bibnamefont{Bari}},
  \bibinfo{author}{\bibfnamefont{B.}~\bibnamefont{De~Maria}},
  \bibinfo{author}{\bibfnamefont{S.}~\bibnamefont{Guzzetti}},
  \bibinfo{author}{\bibfnamefont{R.}~\bibnamefont{Colombo}}, \bibnamefont{and}
  \bibinfo{author}{\bibfnamefont{F.}~\bibnamefont{Raimondi}},
  \bibinfo{journal}{Frontiers in physiology} \textbf{\bibinfo{volume}{6}}
  (\bibinfo{year}{2015}).

\end{thebibliography}


\end{document}